\documentclass[aps,pre,twocolumn,superscriptaddress]{revtex4}

\usepackage{graphicx}
\usepackage{color}

\begin{document}
\title{Properties of pedestrians walking in line: Stepping behavior }
\author{Asja Jeli\'c}
\email[]{asja.jelic@gmail.com}
\affiliation{Laboratory of Theoretical Physics, CNRS (UMR 8627), University Paris-Sud, Batiment 210, F-91405 Orsay Cedex, France}
\affiliation{Istituto Sistemi Complessi (ISC--CNR), Via dei Taurini 19, 00185 Roma, Italy}
\affiliation{Dipartimento di Fisica,``Sapienza'' Universit\'a di Roma, P.le Aldo Moro 2, 00185 Roma, Italy}
\author{C\'ecile Appert-Rolland}
\email[]{cecile.appert-rolland@th.u-psud.fr}
\affiliation{Laboratory of Theoretical Physics, CNRS (UMR 8627), University Paris-Sud, Batiment 210, F-91405 Orsay Cedex, France}
\author{Samuel Lemercier}
\email[]{samuel.lemercier@irisa.fr}
\affiliation{INRIA Rennes - Bretagne Atlantique, Campus de Beaulieu, F-35042 Rennes, France}
\author{Julien Pettr\'e}
\email[]{julien.pettre@irisa.fr}
\affiliation{INRIA Rennes - Bretagne Atlantique, Campus de Beaulieu, F-35042 Rennes, France}

\date{\today}

\begin{abstract}
In human crowds, interactions among individuals give rise to a variety 
of self-organized collective motions that help the group to effectively solve
the problem of coordination.
However, it is still not known exactly how humans adjust their 
behavior locally, nor what are the direct consequences on the emergent organization. 
One of the underlying mechanisms of adjusting individual motions is the 
stepping dynamics.
In this paper, we present first quantitative analysis on the stepping behavior 
in a one-dimensional pedestrian flow studied under controlled laboratory conditions.
We find that the step length is proportional to the velocity of the pedestrian,
and is directly related to the space available in front of him, while the 
variations of the step duration are much smaller.
This is in contrast with locomotion studies performed on
isolated pedestrians and shows that the local density has
a direct influence on the stepping characteristics.
Furthermore, we study the phenomena of synchronization --walking in
lockstep--
and show its dependence on flow densities. 
We show that the synchronization of steps is particularly important at high 
densities, which has direct impact on the studies of optimizing pedestrians
flow in congested situations. 
However, small synchronization and antisynchronization effects are
found also at very low densities, for which no steric constraints exist between 
successive pedestrians, showing the natural tendency to synchronize according 
to perceived visual signals.
\end{abstract}


\maketitle


\section{Introduction}

As in many biological systems, such as fish schools, flocks of birds, or ant colonies,
the dynamics of large pedestrian groups are governed by local interactions
between individuals which give rise to a variety of collective motions occurring
on a macroscopic scale \cite{vicsek_z2012,cividini_a_h2012}. 
Such self-organized behaviors of large pedestrian groups are studied
for practical and safety reasons --improving pedestrians facilities and 
preventing accidents in emergency regimes-- but also as an intriguing 
problem of out-of-equilibrium physics. 

When pedestrians are separated by a small distance,
they cannot walk freely.
It is an open question how they adapt to large
densities. Indeed, it is known that large density
pedestrian flows give rise to increasing fluctuations
in the individual motions, that can eventually lead
to the so-called ``crowd turbulence'' \cite{helbing_j_a2007}
known to be responsible for crowd disasters.
A careful analysis of the behavior of pedestrians
at relatively high densities can give information on
the individual behaviors that could lead to such
transitions.

When pedestrians are very close to each other,
the distance between them can become of the
same order as the longitudinal
displacement due to steps.
Besides, accelerations and decelerations occur
on time scales similar to the stepping period.
Thus, steps cannot be ignored when dealing with
high density flows.
In fact, it was observed in the videos of the experiments
reported in \cite{seyfried2005} that at high
densities, people were walking in lockstep (which we term 
here also ``synchronization'' of steps)
in order to optimize the use of the available space.

Interactions between pedestrians usually take place in
a two-dimensional space and produce velocity changes in direction
and modulus.
However, there are situations where interactions 
are mostly longitudinal, for example,
when people are walking along a narrow corridor.
It is also known that counterflows --pedestrian flows of opposite 
direction-- induce lane formation \cite{kretz2006,moussaid2012}, 
and within these lanes longitudinal interactions should dominate. 
Indeed, some previous experiments have suggested that the adaptations of 
velocity in angle and in modulus could be decoupled to some extent \cite{ondrej2010b}.
This
decoupling has already been used in some
models
\cite{ondrej2010b,moussaid_h_t2011}.

As one-dimensional pedestrian flows involve purely longitudinal interactions
which induce only changes in the velocity modulus \footnote{
Of course a one-dimensional paths may not be straight and
pedestrians may have to change their velocity orientation to follow
the path. However, if direction changes are not too abrupt,
they should have no influence on the velocity modulus.
}, a lot of interest
has been shown recently in studying how pedestrians follow each
other in such settings.
From the point of view of modeling, one-dimensional flows can serve as a 
simple test of a model's ability to produce following behavior.
The experiment is easier to interpret if periodic boundary
conditions are used, i.e. pedestrians walk on a closed line.
Indeed, the transients that may occur when pedestrians enter
or exit an experimental set-up are then avoided.
Besides, the global density is constant in a closed system,
while it is more difficult to control it in an open system.
Such experiments have 
been already conducted in recent years.
Seyfried et al. \cite{seyfried2005,chattaraj_s_c2009,seyfried2010}
have performed experiments with pedestrians following an oval
path.
Similar studies have been reported in references \cite{jezbera2010,yanagisawa_t_n2012}.
These experiments have been performed either by using video analysis 
of the individual trajectories along one straight portion of the set-up 
\cite{seyfried2010}, or by measuring times at which each participant passes
a given measuring point \cite{yanagisawa_t_n2012}. 

We have reported in \cite{jelic2012a} new experiments where
pedestrians were asked to follow
circular paths.
Pedestrians were tracked with a high precision motion capture device.
As a result of being able to cover a larger range of densities than
in previous experiments,
we found that the behavior of a pedestrian following another one
was exhibiting two transitions, when the distance between the pedestrians
was becoming respectively less than $1.1$ and $3$ meters~\cite{jelic2012a,appert-rolland2012}.
Beyond this first result, 
as our experiments provide high precision tracking of all the
individual trajectories during the whole duration of the experiment,
we could access other 
features, either at more macroscopic scales (e.g.\ forming of jams)
or microscopic scales (e.g. stepping dynamics), which were not accessible in previous experiments. 

In this paper, we focus on the stepping dynamics
in the one-dimensional pedestrian flow experiment 
presented in \cite{jelic2012a}. 
We perform the first quantitative analysis of
the stepping dynamics in large pedestrians groups.
We show that the size of steps is directly related to
the space available in front of the pedestrian,
and that the step frequency is far less sensitive
to the local density. 
Furthermore, we examine the effect of the flow densities on 
the synchronization of steps among the consecutive pairs of 
pedestrians. 
We find as expected that a certain amount of synchronisation occurs at
high densities.
However, more surprinsingly, we also find some synchronisation
at lower densities.
Besides, at lower densities we also found occurrence 
of an ``antisynchronization'' phenomenon, i.e. a consecutive
pair of pedestrians can be synchronized in such a way 
that when one of them is stepping with the left leg, the
other is stepping with the right leg, and vice versa.
This natural tendency to synchronize could actually
be exploited to improve the 
flow in a congested situation, for example using
the effect of  rhythm and music 
on the stepping behavior \cite{yanagisawa_t_n2012,styns2007}.

\begin{figure*}[t] 
  \begin{center}
   \includegraphics[height=0.3\textwidth]{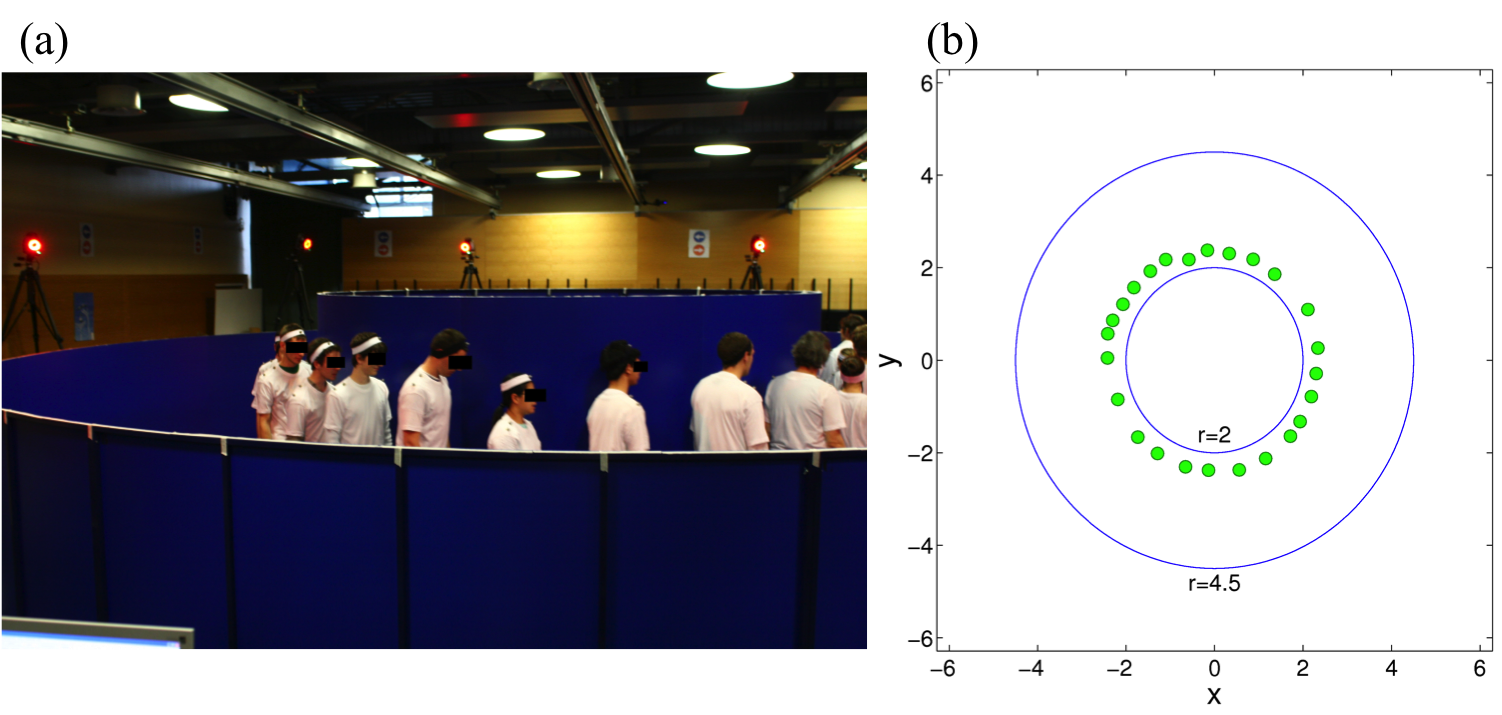}
    \includegraphics[height=0.3\textwidth]{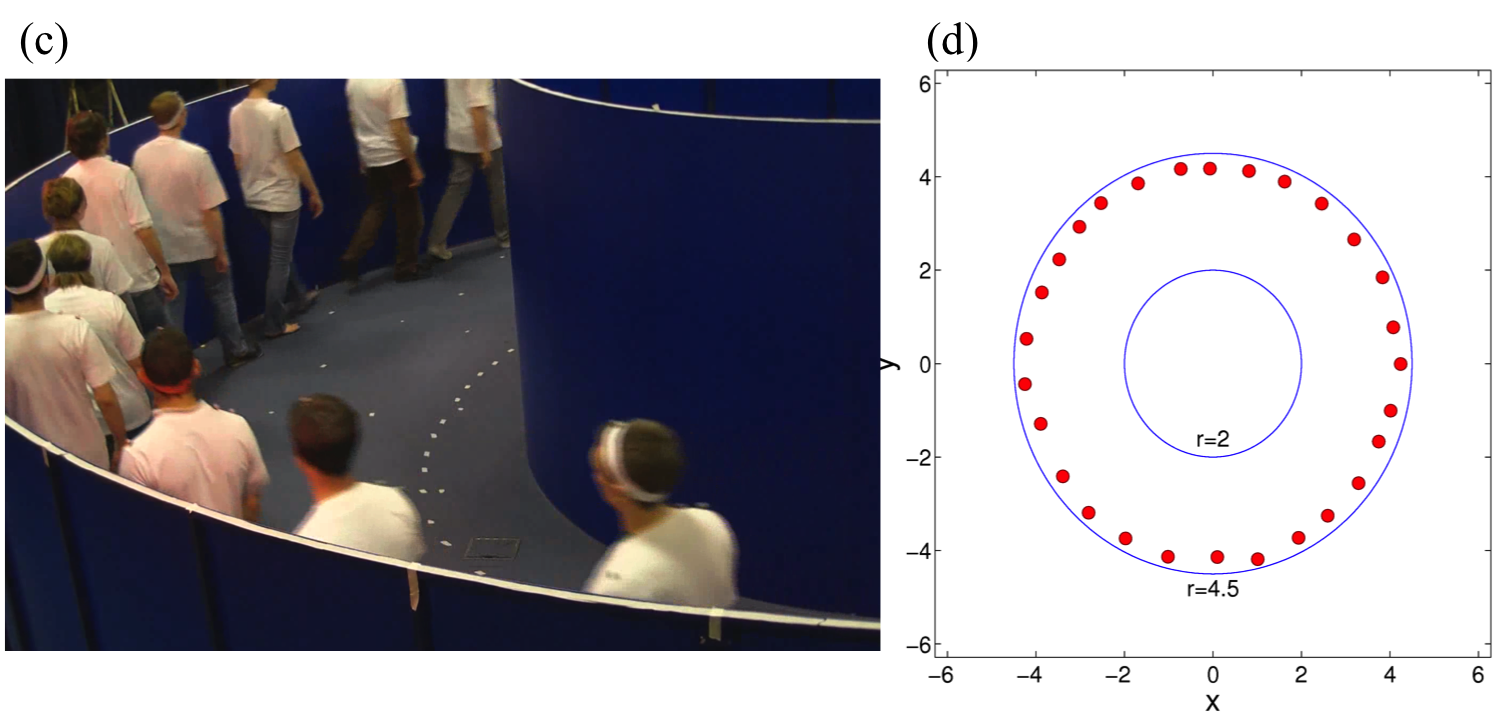}
  \end{center}
   \caption{
   (Color online)
   Experimental set-up: (a) and (c) photo of the experiment and (b)
   and (d)
   top view obtained through {\sc MATLAB} treatment of the data,
   showing a snapshot of the pedestrians positions (small circles).
   The experimental set-up is made of two cylindrical walls
   surrounded by infra-red cameras.
   In the experiment, pedestrians were asked to walk either along
   the inner wall (top images) or along the outer wall (bottom
   images).
  The top figures are taken from Ref.\ \cite{jelic2012a}, which describes 
	  another aspect of the same experiment. }
\label{fig:ring}
\end{figure*}

First we shall summarize the experimental protocol (section
\ref{sect_exp}).
Step measurements are described in section \ref{sect_step_mes}.
In section \ref{sect_law} we present results showing that step 
length and duration obey simple laws.
Section \ref{sect_synchro} is devoted to the study of step synchronization.

\section{The experiment}
\label{sect_exp}

The aim of the experiment was to study the
longitudinal interactions between pedestrians
walking in line, without overpassing each other, along 
a circular path.
While the study of the fundamental diagrams and velocity-spatial 
headway relation was presented in \cite{jelic2012a},
here we focus on the extraction of the stepping
behavior in the various dynamic regimes
(free flow, jammed, etc.). 

As detailed in \cite{jelic2012a},
the experiment was performed inside a ring corridor
formed by inner and outer circular walls of radii
$2$ and $4.5$ m, respectively (see figure \ref{fig:ring}).
Participants
were told to walk in line along either the inner or outer
wall, without passing each other.
As a result, we obtained two
types of pedestrians trajectories:
along the inner circular path the observed average radius of
the trajectories was $2.4$ m, 
and along the outer circular path, the observed radius was $4.1$ m.

Pedestrians were volunteers, unaware of the goal of the 
experiment. In order to capture their most natural (real situation) 
behavior, even though in laboratory conditions, they were 
asked to walk in a ``natural way'', as if they were walking 
alone in the street (and without talking to each other).
Up to 28 pedestrians 
(20 males and 8 females)
were involved in the experiment.
The average global density was varied from $0.31$ to $1.86$
ped/m
\footnote{For bi-dimensional pedestrian crowds, densities are
expressed in ped/m${}^2$. An attempt to make a connection between
one- and bi-dimensional densities was proposed
in~\cite{seyfried2005}.
It is based on the assumption that the lateral
width required by the one-dimensional flow increases with the
velocity.
If we use a transformation such as suggested in~\cite{seyfried2005},
namely 
$\rho_{1D\rightarrow 2D} = \rho_{1D} / (0.46 + 0.2 v)$,
then the estimate for the density range covered by our experiment
would extend from $0.4$ to $3.7$ ped/m${}^2$ (using the
velocity-density relation measured in~\cite{jelic2012a}). But
of course these figures
should be taken only as a rough estimate.
}
by varying both the number
of participants involved, and the length 
of the circular trajectory.

Each participant was equipped with $4$ markers
(one on the left shoulder, two on the right shoulder,
and one on top of head).
Motion was tracked by 12 infra-red cameras (VICON MX-$40$
motion capture system).
The raw data were turned into 3D markers trajectories using
the reconstruction software VICON IQ,
with a frequency of 120 frames per second 
(for more details see \cite{lemercier2011b}).

The trajectories of the markers belonging to the
same pedestrian were aggregated
\footnote{
Some special care was necessary at this stage,
since markers may be temporarily hidden by the walls or
participants' bodies. We kept only data for which
at least two markers per pedestrian were visible.}
in order to give one single three-dimensional trajectory 
for each pedestrian:
its radial, angular and height coordinates are given as a function
of time during the whole duration of experiment.
The corresponding velocities are easily calculated.

\section{Step measurements}
\label{sect_step_mes}

In our previous paper on these experimental data 
\cite{jelic2012a}, we were interested in properties like
fundamental diagrams and velocity-spatial headway relations, 
and therefore we used filtering in order to eliminate the 
oscillations of the trajectories due to 
stepping.
Here, on the contrary, we focus on these oscillations.

\begin{figure}[b] 
  \centering
         \includegraphics[width=0.45\textwidth]{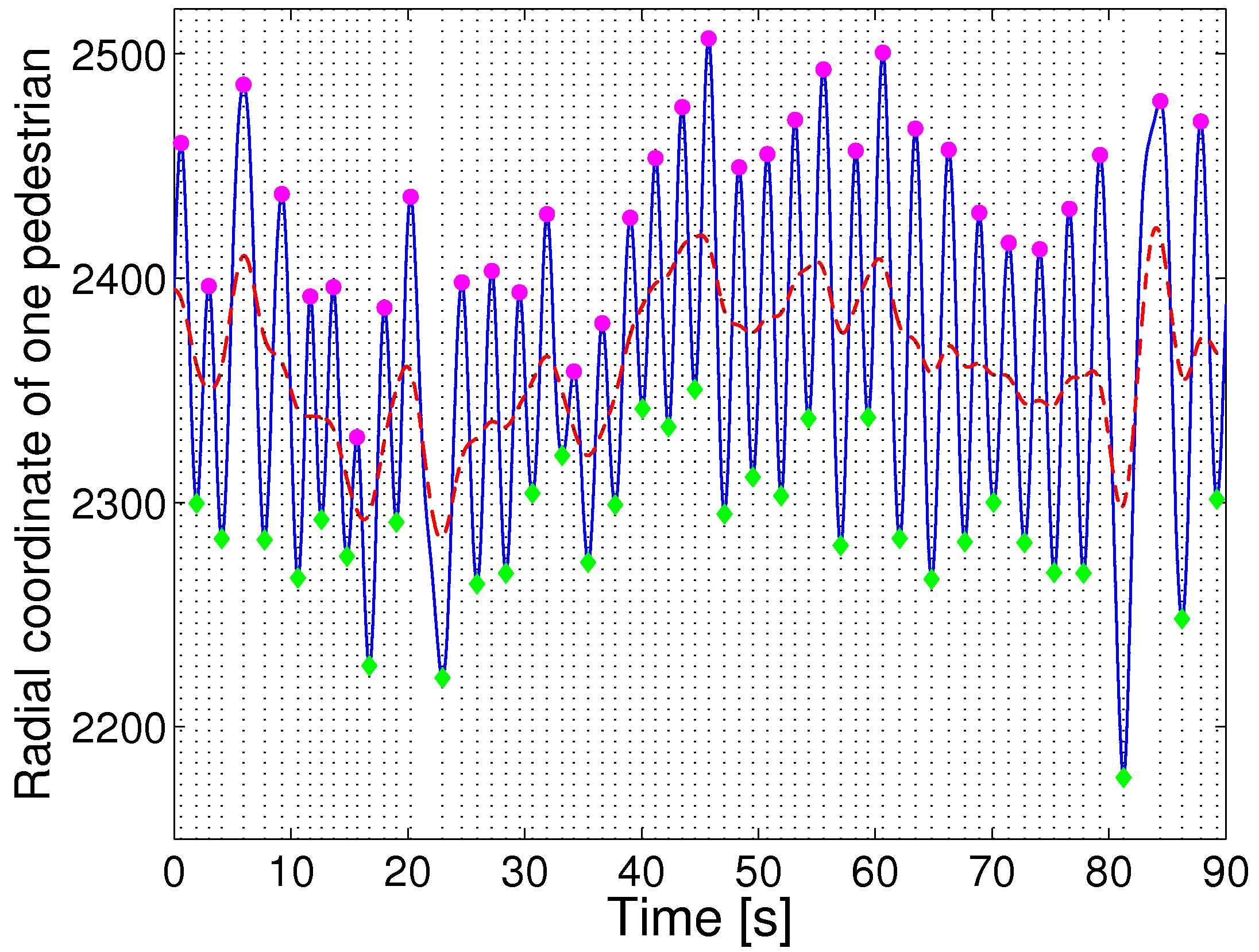}
   \caption{(Color online)
The full (blue) line is the non-filtered radial coordinate 
of one participant (in mm), walking along the inner
circle trajectory of average radius  $2.4$ m, with clearly 
visible oscillations due to 
the consecutive steps. 
The dashed (red) line is the filtered radial
coordinate (we used a 2nd order low pass butterworth filter).
Thin vertical lines ($..$) mark the local extrema --local 
maxima (magenta circles) and minima (green diamonds)-- 
which are used in our definitions of step characteristics. 
}
   \label{fig:steps-determination}
\end{figure}

The whole body of a pedestrian is swaying when the body weight
is shifted from one leg to the other.
This swaying results in oscillations that can be seen
on the three coordinates of a
given pedestrian. However, the angular coordinate oscillations
are entangled
with the average forward motion of pedestrians,
and the height data can be spoiled with
spurious motion of the pedestrian head.
Therefore, we found that
the radial coordinate is the one yielding
the best signal for the detection of the stepping cycles
(see Fig.\ \ref{fig:steps-determination}).
Indeed, stepping induces some lateral body movement clearly
visible on the radial coordinate.
Besides, as pedestrians are walking along circular paths, 
the radial coordinate is decoupled from the forward motion.

The high precision of our experimental measurements 
allowed to extract information on 
the step length, step duration, and also to analyze
the synchronization phenomena between two successive pedestrians.

We define step duration as the time $\Delta t_s$ passed between
two consecutive local extrema in the radial coordinate
(consecutive local minimum and maximum of oscillations).
The step length is then the distance that a participant
has traveled along the circle during this time. 
It is defined as $l_s=\Delta\theta_s\left<R\right>_{s}$,
where $\Delta \theta_s$ is the angle covered during time $\Delta t_s$,
and $\left<R\right>_{s}$ the average of the radius of 
a circular trajectory along which pedestrian is walking
during step duration $\Delta t_s$.

One may argue that, as we do not track directly the foot motion,
the oscillations that we detect on the radial coordinate
could not rigorously coincide with the feet cycles.
However, on average the step length and step duration
values should be the same.

We performed analysis on the set of all steps made
by each participant during all of our $52$ experiments,
each lasting about $1$ minute. 
On average, step duration is estimated to be of the order of $1$ s, 
meaning each participant made approximately $60$ steps during one 
experiment.
We kept only data with at least $2$ visible markers at
each time frame between two steps (two local extrema), 
the rejection ratio being around $10\%$.

\begin{figure}[t] 
  \centering
   \includegraphics[width=0.23\textwidth]{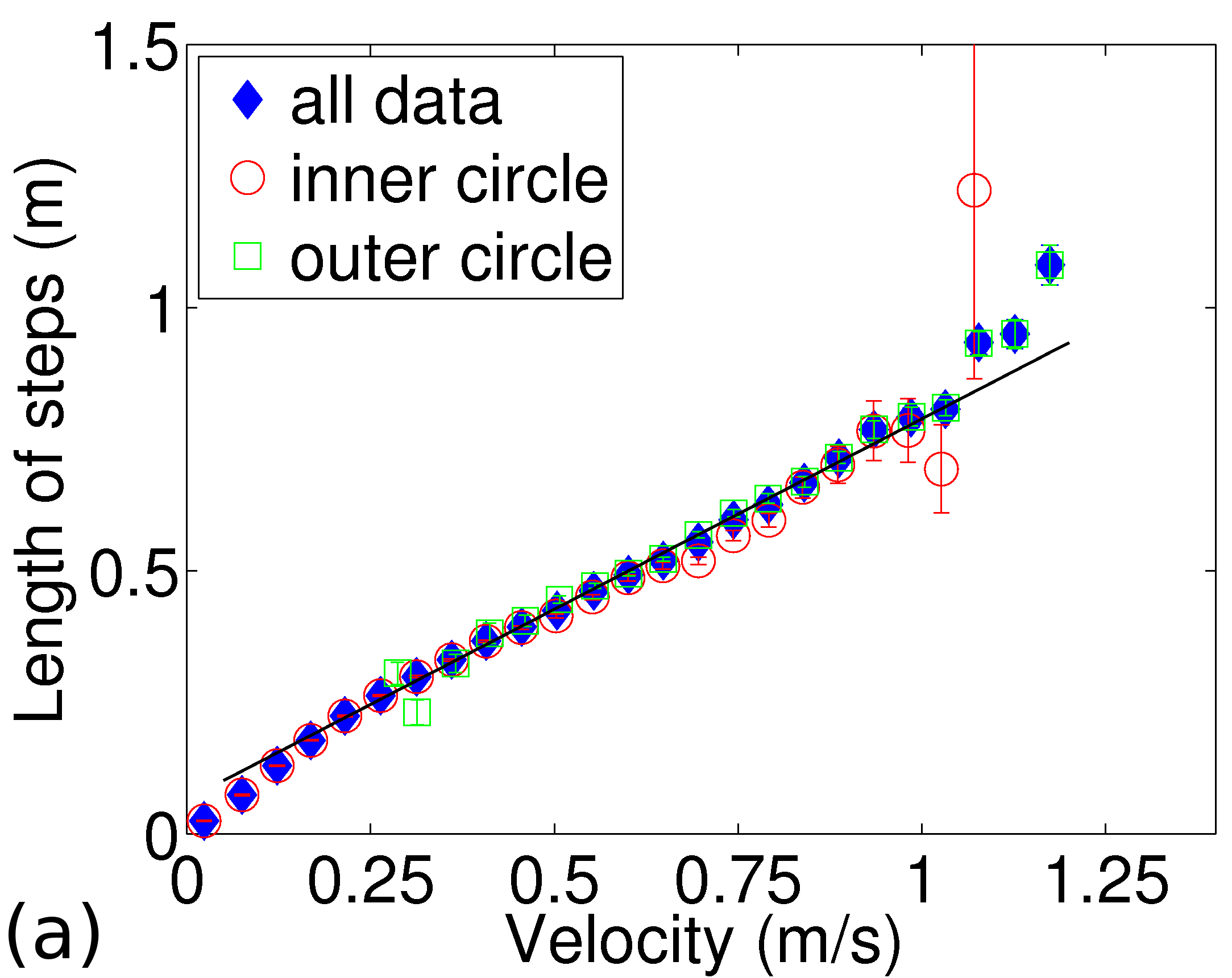}
   \includegraphics[width=0.23\textwidth]{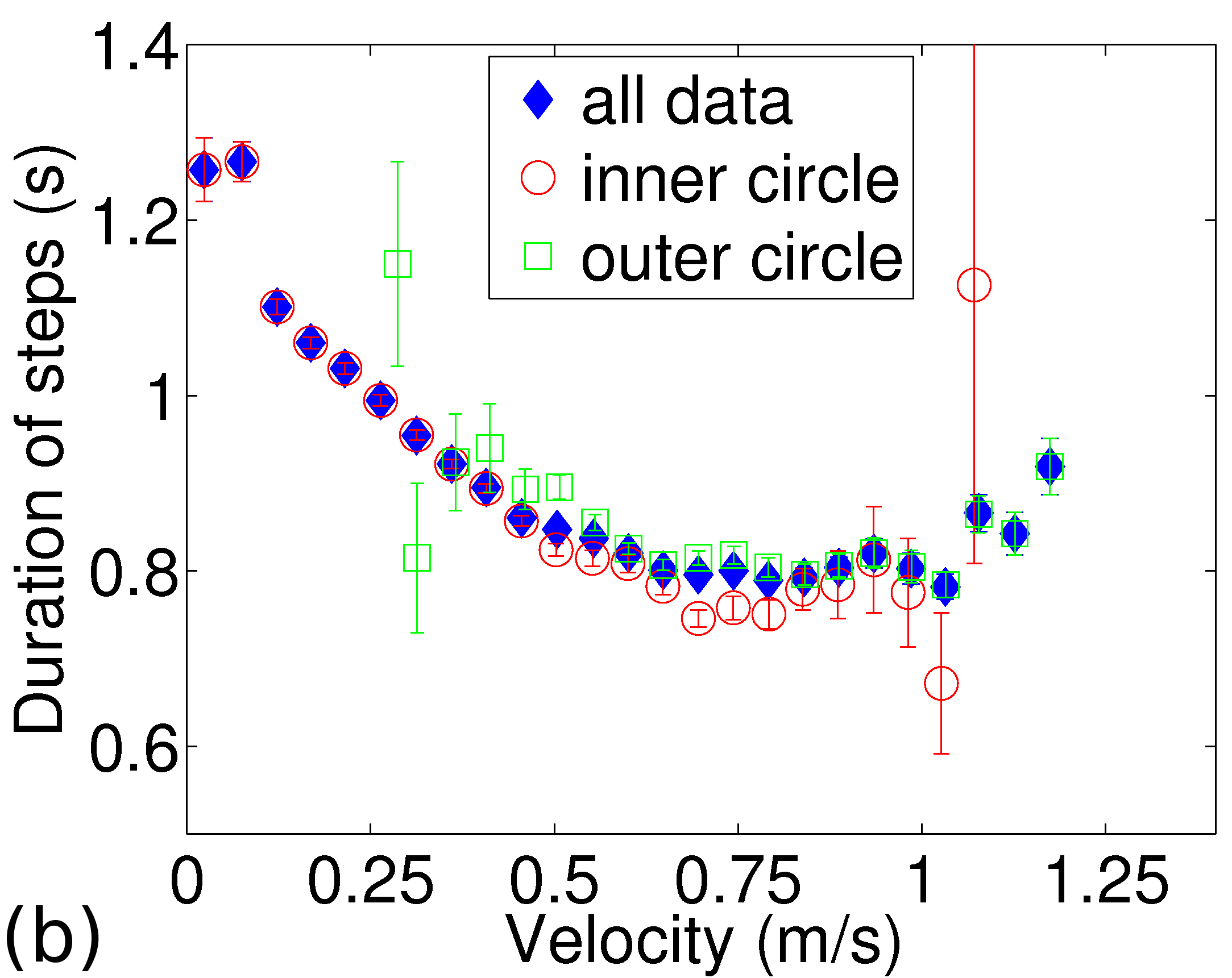}
\\
   \includegraphics[width=0.23\textwidth]{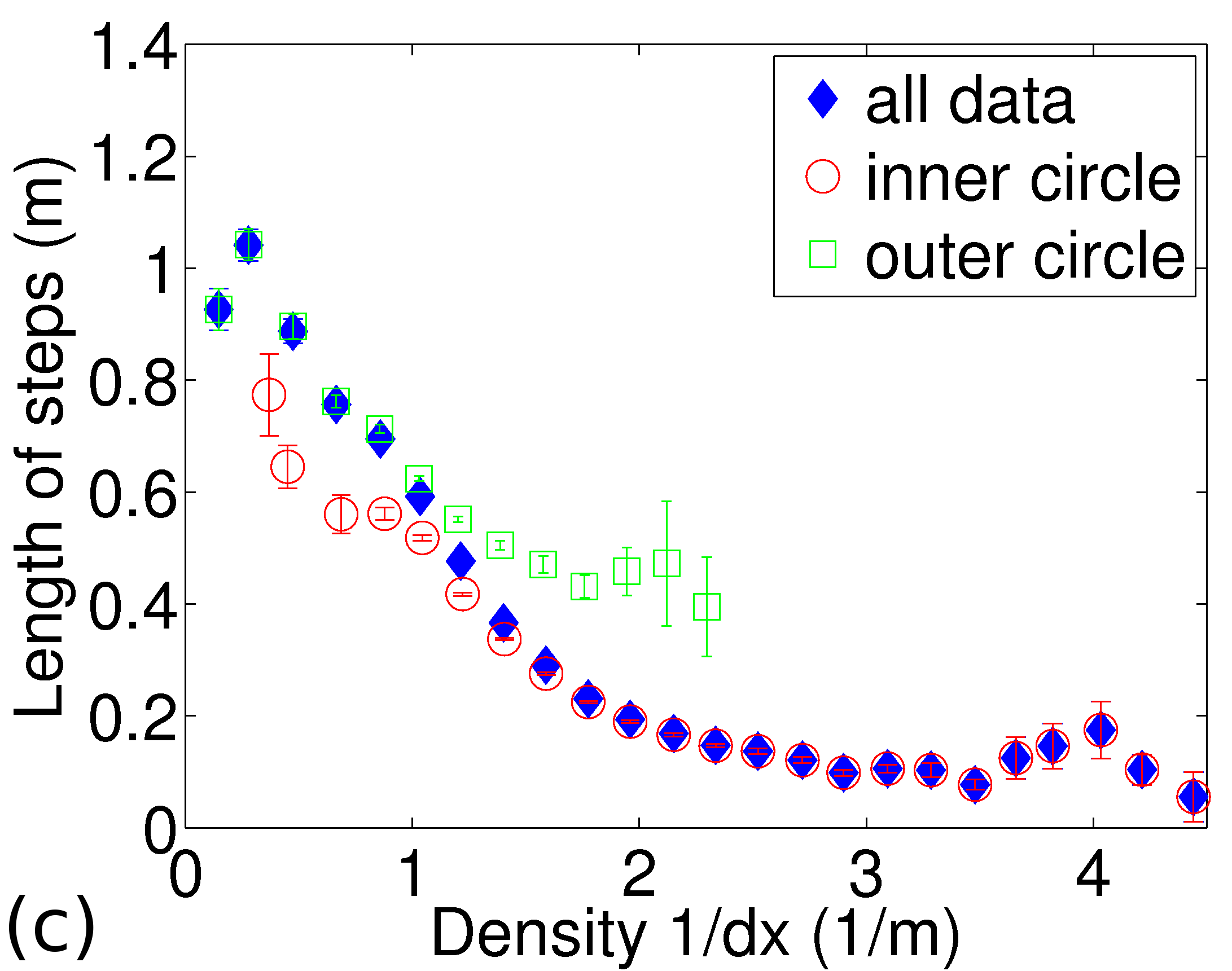}
   \includegraphics[width=0.23\textwidth]{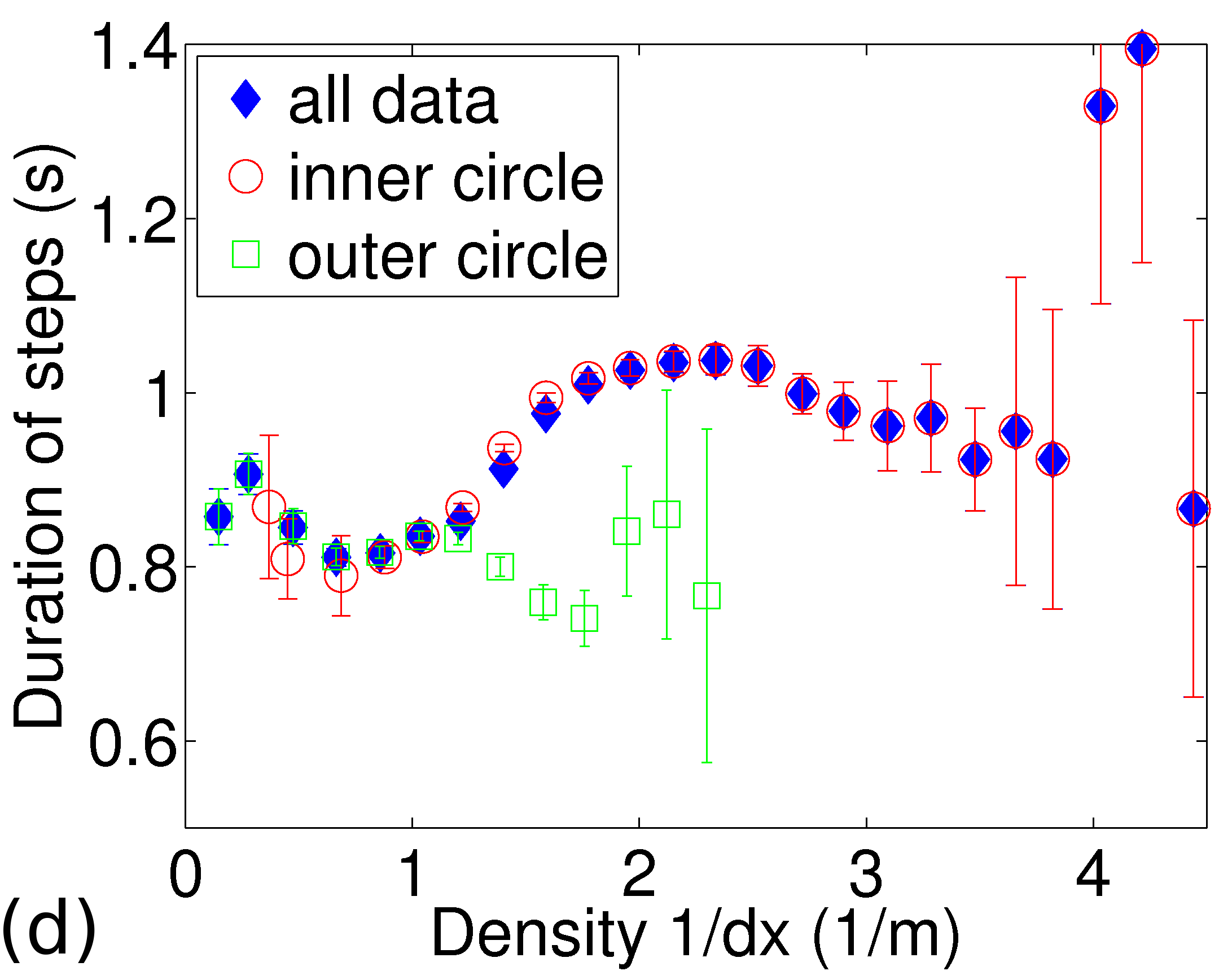}
   \caption{(Color online)
Dependence of the step length (left) and duration (right) as a function of the instantaneous 
velocity (top) and density (bottom).}
   \label{fig:steps_all}
\end{figure}

\section{Stepping laws}
\label{sect_law}

In figure \ref{fig:steps_all} we show the results for the
relationship between
the step length (respectively step duration) and the instantaneous
velocity (top) and density (bottom).
The values of velocity and density were obtained as averages of
the instantaneous velocity and density during the duration of a
given step.
For two successive pedestrians, instantaneous density for the pedestrian
in the back is obtained as the inverse of the distance that is available 
in front of him, i.e.\ to his predecessor.

The most striking feature is that 
the step length is overall proportional to the velocity,
up to velocities very close to zero
(see Fig.\ \ref{fig:steps_all} (a)). Of course,
the step length decreases when the velocity decreases.
Having a vanishing step length for vanishing velocities
means that, within a jammed regime, when pedestrians are 
forced to come almost to a stop, they continue to sway and
shift their body weight from
one leg to the other without moving forward.
A linear fit for velocities between $0.2$ to $1.1$ m/s gives
\begin{equation}
l_s = 0.065m + 0.724 v.
\end{equation}
This differs from the expression given in \cite{weidmann1993}
and used in \cite{seyfried_p_s2010} ($l_s = 0.235 m + 0.302 v$)
for which a residual length of 0.235 m was found at vanishing
velocity. However the details of the measurements were not
given in these references.

As the velocity of a pedestrian is mechanically 
produced by the steps,
we have that
$v=l_s/\Delta t_s$.
As a first approximation, one can state from 
figure \ref{fig:steps_all} (a) that the velocity
is proportional to the step length, and, as a consequence, 
the step duration should be constant.
This is indeed what we find in Fig. \ref{fig:steps_all} (b)
when the velocity is large enough
(larger than $0.6$ m/s): the step duration is then
mostly constant and takes a value around $0.8$ s.
Even for velocities below $0.6$ m/s, the variations
in the step duration are only of 30\%, to be compared
with the 100\% variation in the velocity.
In a more accurate description, this increase
of the step duration when the velocity
of the participants decreases should be taken
into account, yielding a more
complex relationship between velocity and 
step duration.

At first sight, our results could seem
in contradiction with those
obtained in the field of locomotion studies:
Inman's law \cite{inman_r_t1981}
states that the step length \cite{curtis_m2012,boulic_t_t1990}
or step frequency \cite{glardon2005} both vary
as the square root of the velocity.
This law has been indeed verified in
locomotion experiments \cite{glardon2005,dean1965}.
Note that Inman's law requires to normalize data
either by the hip joint height, or total height of
the pedestrian,
an information that is not available in our case,
but that could be measured in future experiments.
However, this renormalization cannot explain the
difference with our result.

In fact, it must be noted
that locomotion experiments are always performed
with {\em isolated} pedestrians. The pedestrian makes
a conscious decision to walk slowly,
and knows that he will
keep this slow pace for a while.
In our experiment, participants walk slowly only
because they are prevented to walk faster by
other participants. Besides, they expect to
be able to walk faster in a near future.
As a consequence of these features, our pedestrians
keep a constant pace (to enable a quick restart),
and rather adopt small step lengths (to comply with
steric constraints).

We have observed that pedestrians
continue to take step even for vanishing velocities.
In order to interpret this result, 
it should be underlined that the vanishing velocities
measured in Fig. \ref{fig:steps_all} (a) are transients:
pedestrians perform one or two steps with vanishing
amplitude and then start again moving forward.
In the case where pedestrians would be standing for
a longer time, one could expect pedestrians to
be more reluctant to
use their energy in swaying when they cannot move forward.
It is also the transient nature of the flow that
could explain that the step duration is bounded
within a 30\% variation:
pedestrians probably do not like
rapid modifications of their stepping pace.

As a conclusion, we observe that in a constrained
environment, pedestrians rather adapt their velocity
through their step length rather than step frequency.
Note that, in Figs.\ \ref{fig:steps_all} (a) and (b),
the data obtained both along the inner and 
outer circle fall on top of each other.
This seems to indicate that there is little influence of
geometry on the stepping behavior.

Let us now consider the dependance of the steps
characteristics with the density. 
The behavior of the step length as a function of
density
(see Fig.\ \ref{fig:steps_all} (c)),
is exactly of the same form as the fundamental diagram
found in \cite{jelic2012a}.
Indeed, this is a direct consequence of the above 
result that the step length is being proportional to velocity.
As the average velocity converges towards a finite
non zero value when the density becomes large,
the average step length saturates around
$0.1m$ at large densities.
By contrast, the step duration changes much less
within different walking regimes.
The step duration varies only for around $20\%$
as a function of density
(see Fig.\ \ref{fig:steps_all} (d)).
The saturation of step frequency at low densities (or high
velocities)
could indicate that increasing the step
frequency beyond a certain value is not comfortable for pedestrians.

We had previously found in \cite{jelic2012a} that for
local density fluctuations far away from the average global
density the velocity could be quite different
from the average behavior. We recover here 
(Figs.\ \ref{fig:steps_all} (c) and (d)) these
atypical behaviors. Indeed, global densities
in experiments performed on the outer (inner) circle 
are always below (above) $1$ ped/m,
and thus the tails ('inner and outer circle data')
diverging from the average behavior ('all data')
in Figs.\ \ref{fig:steps_all} (c) and (d)
correspond to large fluctuations.
Still, for the mean behavior, no discontinuity
is seen when going from the inner to the outer
circle data.

\section{Step synchronization}
\label{sect_synchro}

It was already noticed in \cite{seyfried2005}
that at high densities,
as pedestrians do not have much space to walk,
they tend to synchronize their steps, so as
to squeeze the front leg into the hole left
by the front leg of the predecessor. The authors
refer to this phenomenon as walking in lockstep.
In the following, we want to see whether this
tendency is confirmed in our experiments,
and to quantify this effect.

First we have to define synchronization.
Full synchronization is obtained when two successive
pedestrians walk with the same step frequency and
in phase.

If all pedestrians were walking very regularly,
so that their radial coordinate would be a
perfectly periodic signal 
with the same frequency for all pedestrians,
it would be quite easy to measure the phase
between two successive pedestrians.
However, in the experiments, and especially at high densities,
the shape of the oscillations
and the oscillation frequencies
can vary from one pedestrian to another, and for the same
pedestrian from time to time.

\begin{figure}[t] 
  \centering
     \includegraphics[width=0.23\textwidth]{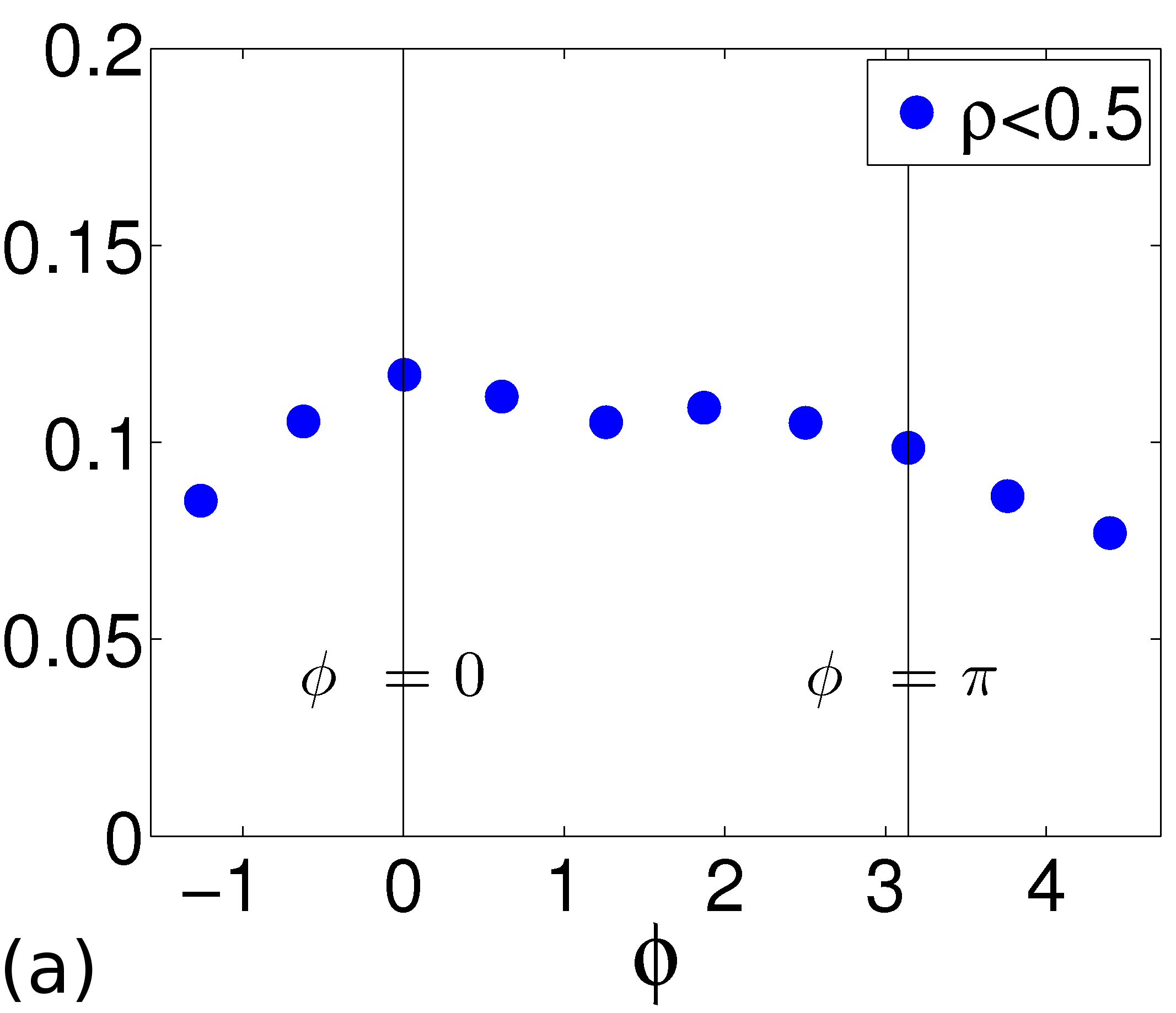}
      \includegraphics[width=0.23\textwidth]{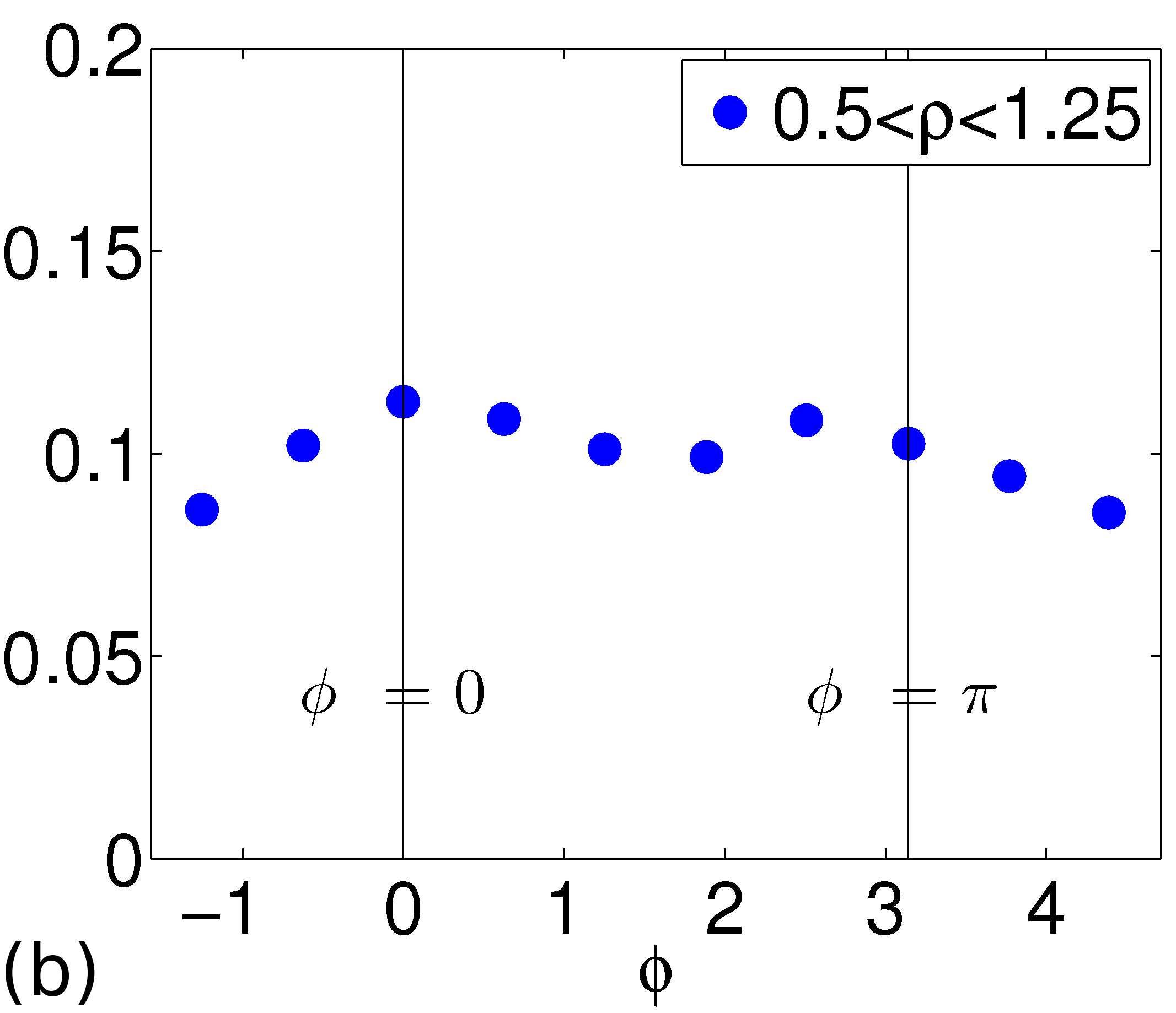}
   \includegraphics[width=0.23\textwidth]{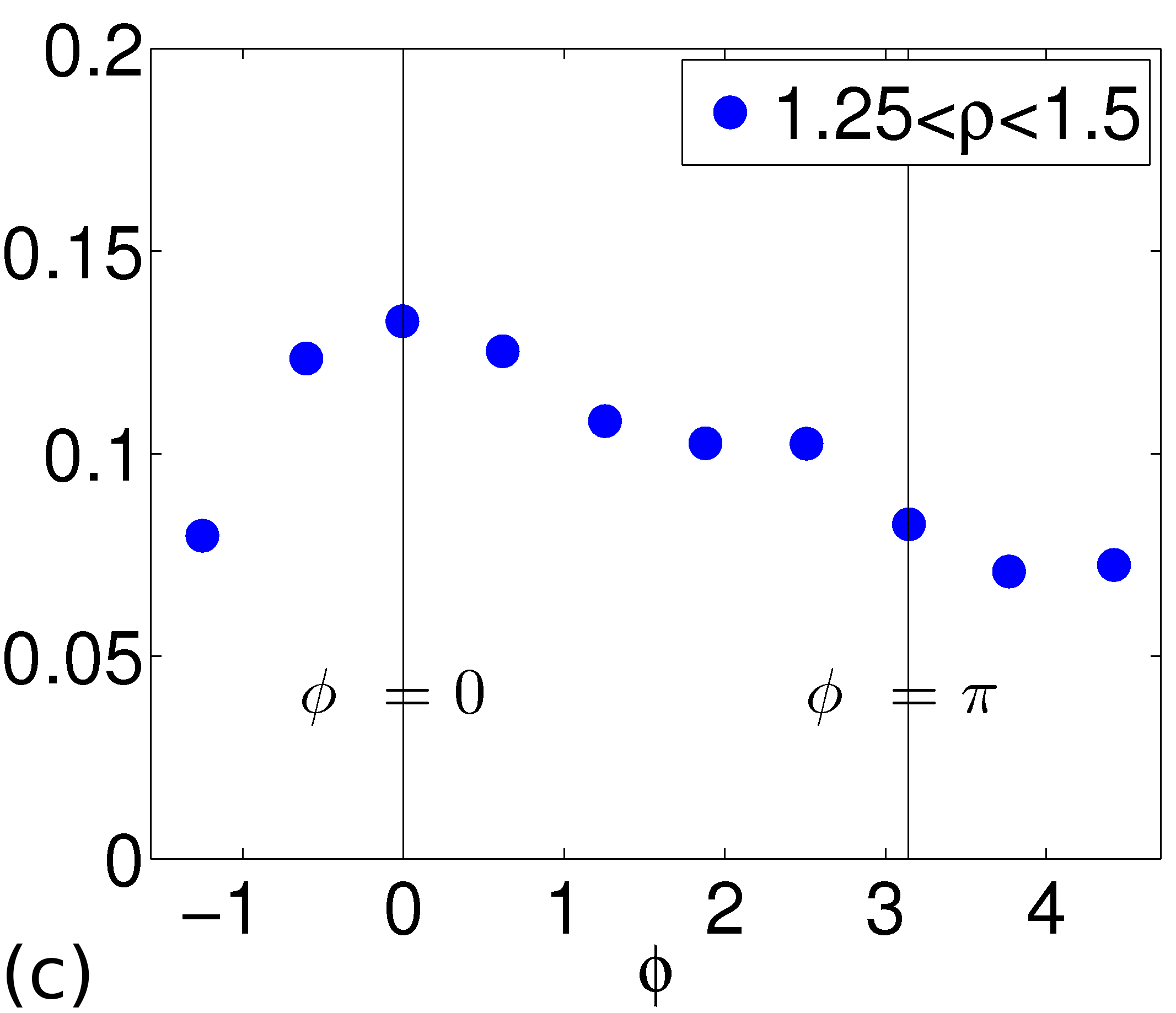}
   \includegraphics[width=0.23\textwidth]{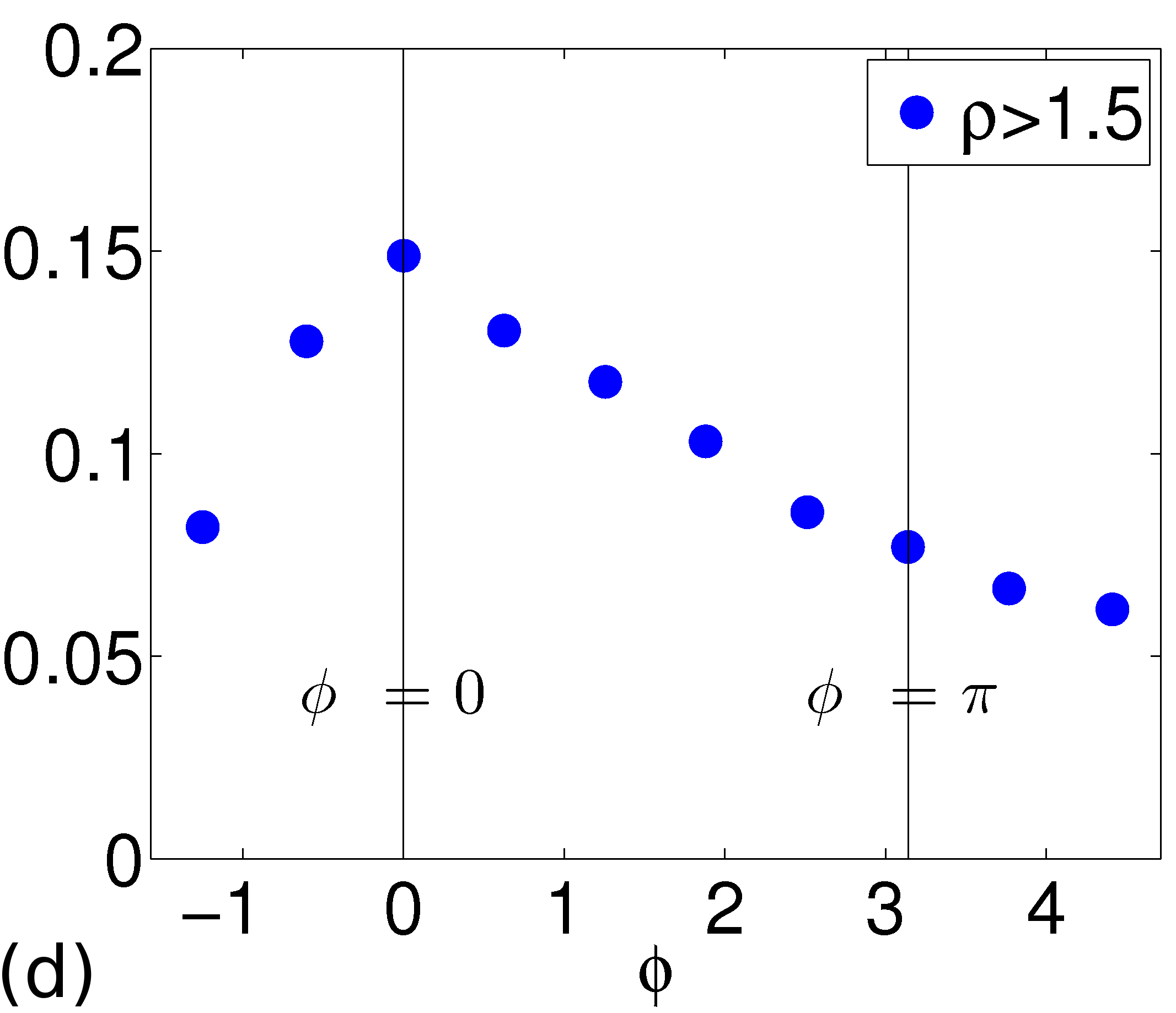}  
 \caption{(Color online)
   Normalized distributions of $\phi$, for various density ranges.
   Synchronization corresponds to $\phi=0$. }
   \label{fig_phi}
\end{figure}

Thus, our measurements included two stages that
we will detail below.
In a first stage, we have analyzed the data
to select pairs of successive pedestrians for which
the stepping frequencies were not too different.
Then, for this subset, we have measured the
phase shift between close minima (maxima) of the stepping 
cycles of the two successive pedestrians.
More precisely, for each detected 
minimum (maximum)
on the radial coordinate of the predecessor 
(occurring at time $t_0$) our analysis consists of the 
following steps:
\begin{itemize}
\item 
We measure the 'individual and local' frequency
of the leader
over 3 periods, defined from the two maxima (minima)
just before and after the given minimum (maximum).
Let $T$ be the average period over these three
cycles.
\item We determine whether there is a minimum (maximum)
in the radial coordinate of the follower within
the time range $[ t_0 - T/4, t_0 + 3 T/4]$.
This minimum occurs at time $t_1$.
\item We evaluate the 'individual and local' frequency
$1/T^\prime$ for the follower, exactly as it was done
for the leader.
\item If $T^\prime$ and $T$ differ less than $25\%$,
we select the current steps of this pair of pedestrians.
The rejection ratio due to too large difference in 
frequencies is up to $20\%$.
This underlines that at least $20\%$ of the pedestrians
are not synchronized with their leader - as synchronization
requires first to have the same frequency.
\item Then we measure the phase
\begin{equation}
\phi = 2\pi (t_1-t_0)/T.
\end{equation}
In a similar way, we also measure the phase $\psi$ separating 
a minimum (maximum) in the radial coordinate of the leader, 
from a maximum (minimum) in the radial coordinate of the follower.
In this way, we expect to obtain more precise measurements
for the antisynchronization phenomena.
\item Finally, we measure the instantaneous density. 
It is defined as the inverse of the distance between the 
centers of mass of the two pedestrians. As this distance 
oscillates with the steps, we found more relevant to evaluate
it on the filtered data. However, the results presented in 
Fig.\ \ref{fig_phi} are similar when non-filtered data are used.
\end{itemize}

Figure \ref{fig_phi} shows the normalized histograms of $\phi$
obtained for various local density ranges.
At large densities (beyond $1.25$ ped/m), 
we observe a peak around phase $\phi=0$, that clearly
indicates existence of the synchronization phenomenon.
This must correspond to the pedestrians walking in lockstep,
as for these densities the steric constraints become
important.

When the density is lower, the peak around zero
is still observed, though it is smaller than for 
higher densities.
Surprisingly, another peak appears around $\phi=\pi$.
This second peak corresponds to antisynchronization,
i.e.\ walking in phase with the opposite legs.
When antisynchronization occurs at high densities,
we could expect that pedestrians would be located at 
different distances from the wall, so that the left leg
of one pedestrian is more or less aligned with the right 
leg of the other.
However, we did not observe any visible effect
of this type in the data. 
Besides, antisynchronization mostly disappears when the 
density is large, while it survives at densities as large 
as $\rho<0.5$ ped/m, for which there are clearly no steric 
constraints between the pedestrians.

\begin{figure}[t] 
  \centering
     \includegraphics[width=0.23\textwidth]{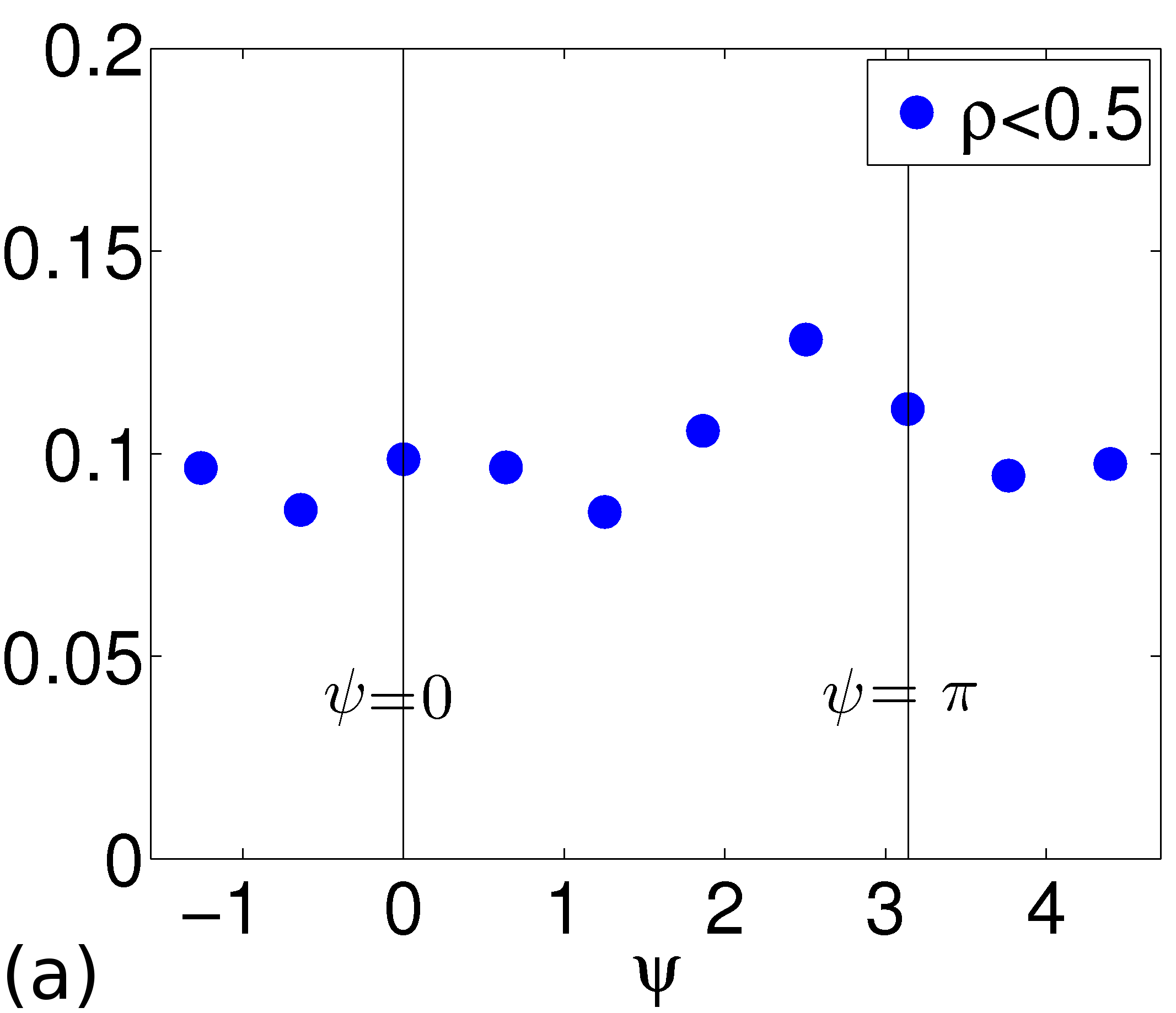}
      \includegraphics[width=0.23\textwidth]{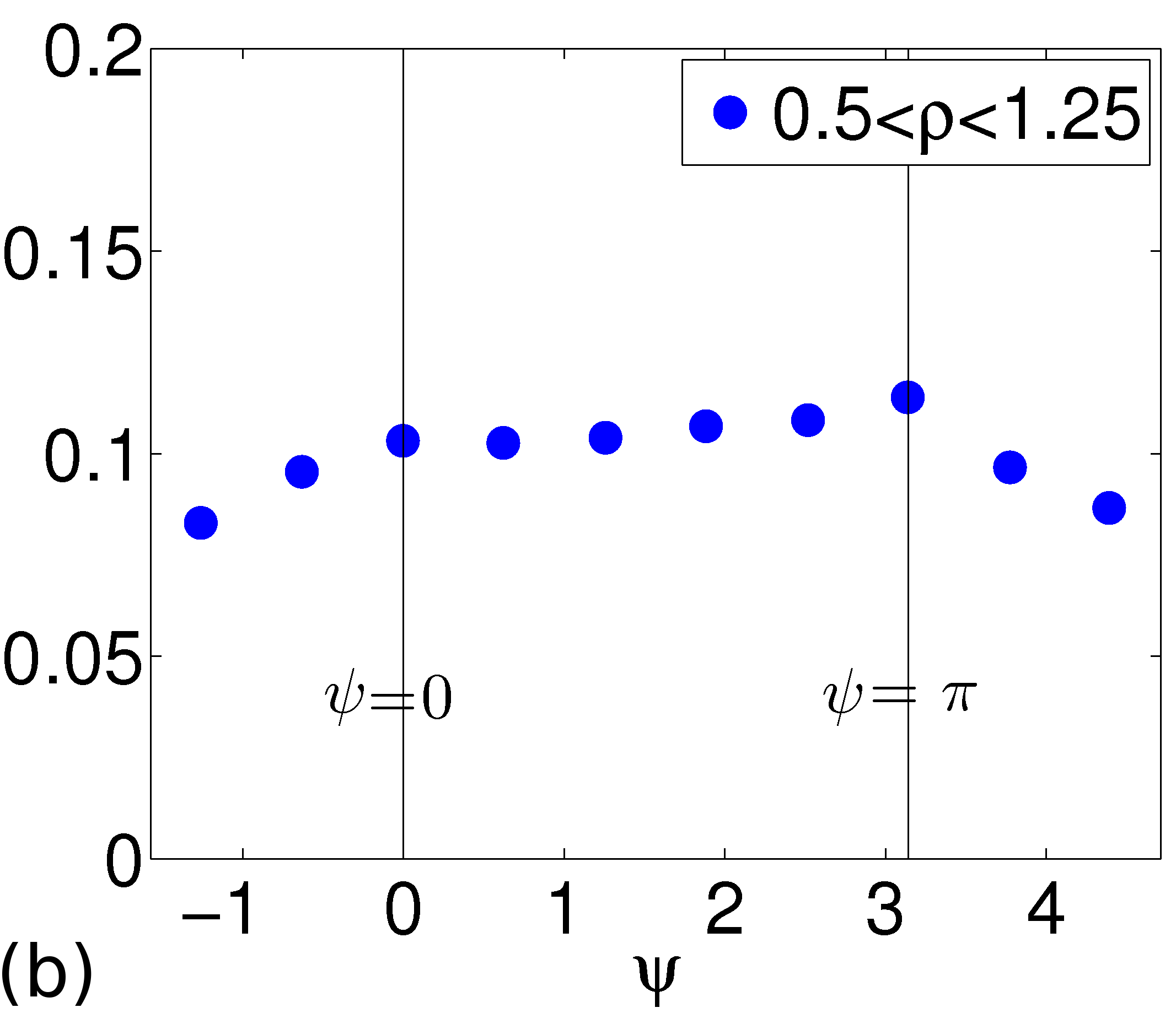}
   \includegraphics[width=0.23\textwidth]{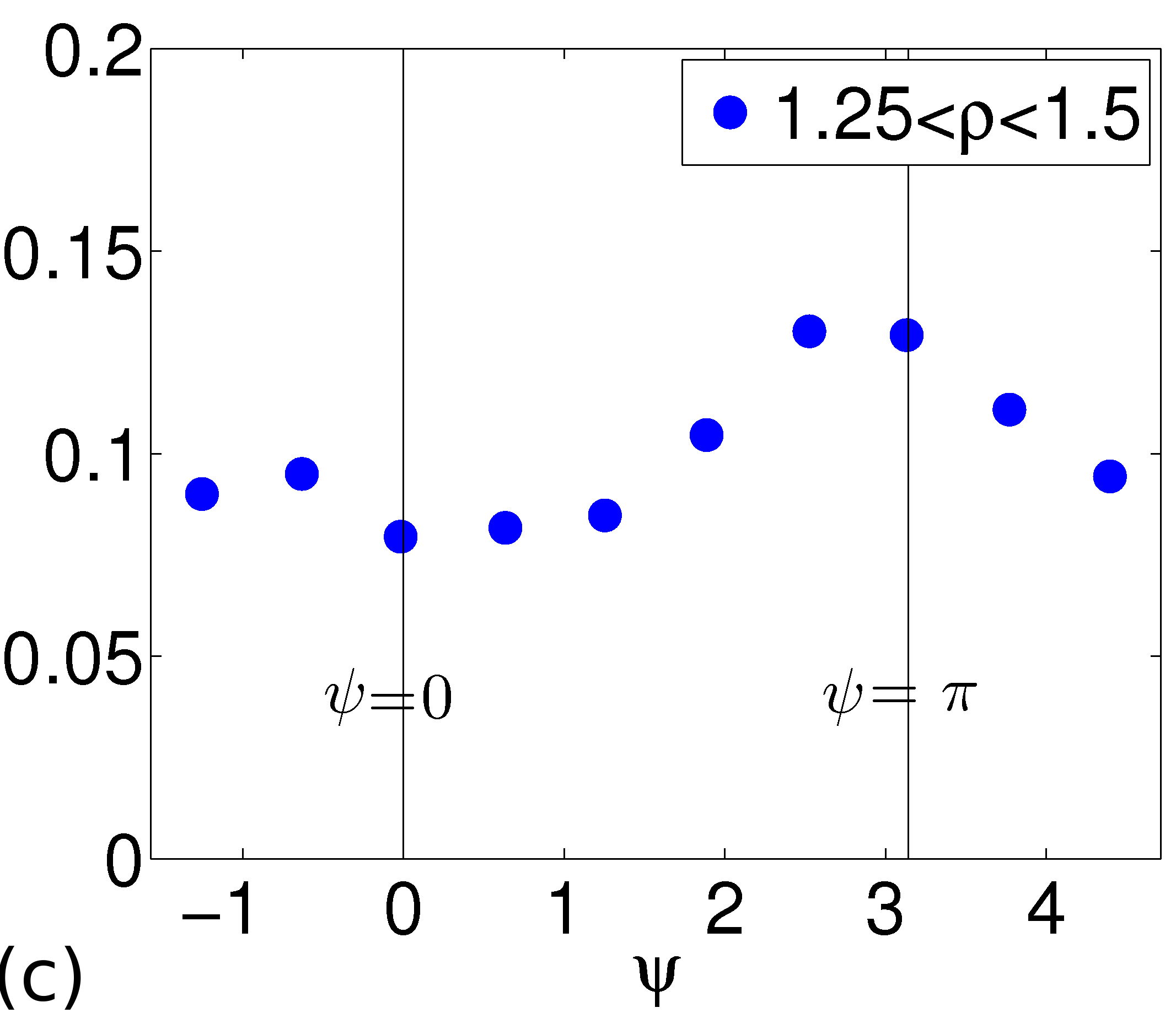}
   \includegraphics[width=0.23\textwidth]{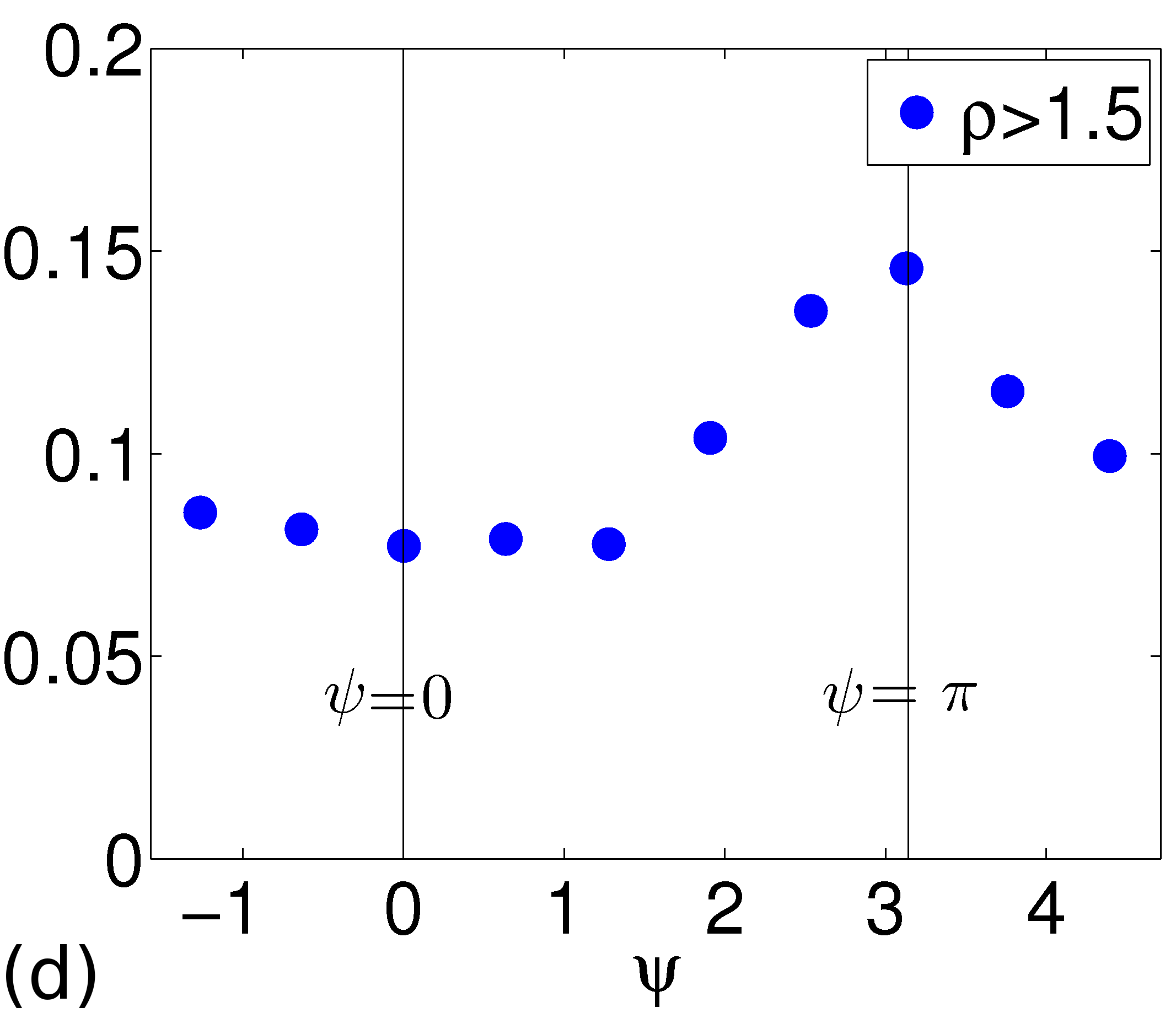}
   \caption{(Color online)
   Normalized distributions of $\psi$, for various density ranges.
   Antisynchronization corresponds to $\psi=0$.
 }
   \label{fig_psi}
\end{figure}

As the pedestrians may not have exactly the same
frequency, in  Fig.\ \ref{fig_psi} we checked if the
antisynchronization is also visible when we measure 
the phase shift $\psi$ between the local extrema of the 
opposite kind (minimum and maximum) in the radial coordinate.
This corresponds to the steps made by the left leg of one 
and the right leg of the other of the two consecutive 
pedestrians.
In this case, antisynchronization should appear as
a peak around zero -- and there is indeed a second peak
located around $\psi=0$ for the lowest densities
as seen in Fig.\ \ref{fig_psi}.
On the other hand, synchronization can be seen again, but 
this time as a peak around $\psi=\pi$.

These results, showing both synchronization and antisynchronization
at low enough densities where the pedestrians are not
bound by the steric constraints, suggest that pedestrians are sensitive
to the stepping oscillations that they perceive visually
when watching their predecessor, and that they naturally
synchronize.
It is still an open question to determine precisely to which visual
signal pedestrians are most sensitive.
Rather than lateral or vertical oscillations of the body,
pedestrians probably perceive more easily the motion of arms and legs
\cite{curtis}.

\section{Discussion and conclusion}
\label{sect_conclusion}

In this paper, we have presented new experimental
results about steps characteristics of pedestrians 
following each other along a one-dimensional trajectory.
We have obtained for a large range of velocities
several simple laws for step length and duration, 
namely that step length is proportional to velocity,
while variations in step duration are much smaller.

This result is in contrast with the hypothesis
used in \cite{johansson2009} that at high densities,
when it is no longer possible to take normal steps,
pedestrians would rather completely stop until they
gain enough space to make a step.
Indeed our observation is that pedestrians
are not reluctant to take very short steps
and continue to 
shift their body weight from one foot to the other, even
when they can almost not move forward, as this is the case
at very high densities.

Our results also highlight that the stepping
behaviors in a crowd can be quite different from
those measured in locomotion experiments
with isolated pedestrians \cite{inman_r_t1981},
though for the same range of walking speeds.

This raises some questions that would be
interesting to tackle in the future.
Indeed, several effects can be responsible
for the change of walking behavior in a crowded
environment. 
Walking very close to other pedestrians
induces physical constraints that obviously
have to be taken into account in the stepping
strategy.
However, at high densities, another effect
comes from the presence of stop-and-go waves:
pedestrians may have different steps characteristics
depending on whether they are accelerating,
decelerating, or walking at constant pace.
Even the anticipation that the pedestrian
will have to accelerate in the near future
could modify his behavior.
It would be interesting to design new experiments
to discriminate between these various effects,
taking also into account the relative height
of interacting pedestrians, and distinguishing
the inter- and intra-pedestrian variations.

Another question that we have addressed in this paper
is whether pedestrians walk in lockstep at
high densities \cite{seyfried2005}.
Indeed, at high densities, beyond $1.25$ ped/m,
we have observed some synchronization between
the step cycles of successive pedestrians.
This can be easily explained by the strong
steric constraints that occur at these densities.
Surprisingly, synchronization is still observed
--though less frequently-- at much lower densities.
It seems that even when pedestrians are more than
$2$ m apart, they still have the tendency
to synchronize their rhythm,
probably as a consequence of the visual
stimulus given by the pedestrian ahead.
Besides, in the absence of steric
constraint, we observed that synchronization and
antisynchronization are both observed
at such low densities.

Interest in the synchronization phenomena
also stems
from the observations that music can induce particular 
stepping behaviors.
In \cite{styns2007}, experiments in which pedestrians were 
asked to synchronize their steps with the indicated  rhythm
were reported. 
It was shown that it was more efficient to indicate
the  rhythm with music than with a simple metronome 
\cite{styns2007}.
In Ref.\ \cite{yanagisawa_t_n2012}, it was found
in an experiment
that when pedestrians walking in line are asked
to walk with a rhythm (given by a metronome) slower
than the natural pace of
pedestrians, the flow in the congested regime is improved,
as a result of the synchronization of steps.
It would be interesting to investigate this further,
and in particular to determine 
the relation between the improvement of the flow and the 
fraction of the pedestrians `synchronized' with the rhythm.

There would be a practical interest in knowing
whether such
synchronization would occur when the music is just used
as a background, without any special assignment, and 
also what would be the consequence
on the macroscopic characteristics
of the flow. 
Further investigations are needed.
It would be necessary to perform new
experiments with tracking methods such as the one described
in this paper, to measure in particular the amount of
step synchronization
between pairs of successive pedestrians.
If it was shown that a musical background can improve
the flow, this could be used in particular as a strategy
to improve evacuation.

\begin{acknowledgments}

This work has been performed within the {\sc PEDIGREE} project,
financed by the French ANR (Contract No. ANR-08-SYSC-015-01), 
and involves four research teams in Rennes (INRIA), Toulouse 
(IMT, CRCA), and Orsay (LPT).
Experiments were organized and realized by the {\sc PEDIGREE}
partnership \cite{pedigree_info} at University Rennes 1, with the help of
the laboratory M2S from Rennes 2. We are in particular
grateful to Armel Cr\'etual, Richard Kulpa, Antoine Marin, and
Anne-H\'el\`ene Olivier for their help during the experiments.
A.J.\ acknowledges support from the RTRA Triangle de la
physique (Project 2011-033T).
We thank Sean Curtis and Ronan Boulic for helpful discussions
about locomotion results.
We are especially grateful to S. Curtis for his careful
reading of our paper and his useful remarks.

\end{acknowledgments}

\bibliographystyle{unsrt}

\end{document}